\documentclass{iaus}
\usepackage{graphicx}

\title[VAO Transient Facility] 
{The VAO Transient Facility}

\author[Graham, Djorgovski, Drake, Mahabal, Williams \& Seaman]   
{Matthew J. Graham$^1$, S. G. Djorgovski$^{1,3}$, Andrew Drake$^1$, Ashish Mahabal$^1$, Roy Williams$^1$
 \and Rob Seaman$^2$}

\affiliation{$^1$California Institute of Technology, Pasadena CA USA \\
email: {\tt mjg, ajd, ashish, roy@caltech.edu, george@astro.caltech.edu} \\[\affilskip]
$^2$National Optical Astronomical Observatory, Tucson AZ USA \\
email: {\tt seaman@noao.edu}\\[\affilskip]
$^3$ Distinguished Visiting Professor, King Abdulaziz Univ., Jeddah, Saudi Arabia}

\pubyear{2012}
\volume{285}  
\pagerange{119--126}
\jname{New Horizons in Time Domain Astronomy}
\editors{R.E.M. Griffin, R.J. Hanisch \& R. Seaman, eds.}
\begin{document}

\maketitle

\begin{abstract}
The time domain community wants robust and reliable tools to enable 
production of and subscription to community-endorsed event notification 
packets (VOEvent). The VAO Transient Facility (VTF) is being designed to 
be the premier brokering service for the community, both collecting and 
disseminating observations about time-critical astronomical transients but 
also supporting annotations and the application of intelligent 
machine-learning to those observations. This distinguishes two types of 
activity associated with the facility: core infrastructure and user services. 
In this paper, we will review the prior art in both areas and describe the 
planned capabilities of the VTF. In particular, we will focus on 
scalability and quality-of-service issues required by the next generation of 
sky surveys, such as LSST and SKA.
\end{abstract}

\firstsection 
\section{Introduction}


The endorsement of the Large Synoptic Survey Telescope (LSST) in the National Research CouncilÕs Decadal Survey of Astronomy and Astrophysics (\cite[Decadal Survey 2010]{decadal}) identified transient astronomy as a vital area of astronomical research for the next decade and beyond. With its stated goals to enable and facilitate science, the VAO is well-placed to deliver the tools and services required by the community to exploit this new domain. Yet this is not a trivial undertaking. The transient astronomy community is a broad church -- a unique burgeoning multi-wavelength group, consisting of both professional and non-professional astronomers. Fortunately, it is already VO cognizant with $\sim$47000 VOEvents having been sent at the time of writing (including $\sim$26000 from the Catalina Real-Time Transient Survey (\cite{crts})). However, it faces daunting data challenges with LSST projecting event production rates of between $\sim$10$^5$ and $\sim$10$^7$ notifications per night and studies showing SKA capable of detecting 1 core-collapse supernovae per second over the whole sky originating somewhere in the redshift range $0 < z < 5$.

Over the past 6 years, two NSF-funded projects -- VOEventNet and SkyAlert -- have already contributed many aspects of the current event infrastructure. The time is now ripe to define the next-generation service -- the VAO Transient Facility -- that will continue this work and put it onto a sound operational footing, capable of addressing the needs of the community. This paper describes a review study which we have carried out concerning the development of such a facility within the VAO built upon the existing SkyAlert product.  We have identified the short- to mid-term requirements of the community and a program to transition the prototype SkyAlert to a production-level service with all that that entails. 

\section{Community engagement}

A VAO Transient Facility aims to provide the transient astronomy community with robust and reliable tools to produce event notifications, to subscribe to event notifications, and to leverage VAO capabilities for data discovery/analysis.
Responses were solicited from a cross-section of the community -- LSST, LOFAR, GAIA, SKA Pathfinders (MeerKAT, ASKAP), AAVSO, Zooniverse, and LIGO -- to identify specific issues within these areas that require attention and to assist in planning and determining priorities. Specific questions were asked about: mode and rates of event delivery, brokering requirements including levels of service and functionality, current standards and infrastructure, repository holdings and performance, and other areas of potential interest such as EPO opportunities. The responses from the community can be broadly grouped under two categories: simplicity and programmatic access. 

It is all too easy to try and make everything as feature-rich as possible to try and meet all possible use-cases. However, there is a strong desire to keep things manageable. Event notifications should just report the basics of a transient event. References can always be used to link to extra data, e.g., images, light curves, etc., but the current version of the VOEvent standard (2.0, \cite{seaman11}) is perfectly adequate. The ability to retrieve stored events based solely on their sky position (RA, Dec) and time meets most anticipated queries. At this stage, querying against any aspect of the full VOEvent data model is not required. There are also concerns whether the existing infrastructure will be able to scale to the event rates of some projects but the existing protocols (Jabber/XMPP and TCPV) seem just fine.  

A browser interface is great for trying out things and getting a feel for what something does but, for operational usage, having an API to code against is much more preferable.  In particular, APIs for the following were requested so that 
it should be possible to define subscription filters using an expression syntax to limit the events received to those of interest. Something akin to the current Python-based syntax used by SkyAlert would be broadly acceptable. Users would also like to be able to provide specific pieces of code (ÔpluginsÕ) to be executed when matching events are received by the broker. There might add extra information to an event report before it is passed on to the subscriber.

\section{SkyAlert}

SkyAlert (http://www.skyalert.org) is an excellent prototype of the type of requested broker service. Instances have been deployed by various projects around the world and it has a strong user base ($\sim$300 subscribers). The VAO has its own set of requirements that products and services must comply with, covering development aspects such as version control and testing as well as operational and user support issues. A review of SkyAlert has been carried out to ensure that it can be integrated into the VAO with minimal effort and also to identify potential problems that we can address, i.e., points of failure and bottlenecks. Two main activities have been identified as being required: the development of tests and documentation. Additional tasks are then required for a production-level transient facility, such as addressing scalability issues --  migrating the code base away from a single server instance -- and potential integration with VAO security mechanisms.

\section{Event infrastructure}

\begin{figure}[h]
\begin{center}
\includegraphics[width=4.55in]{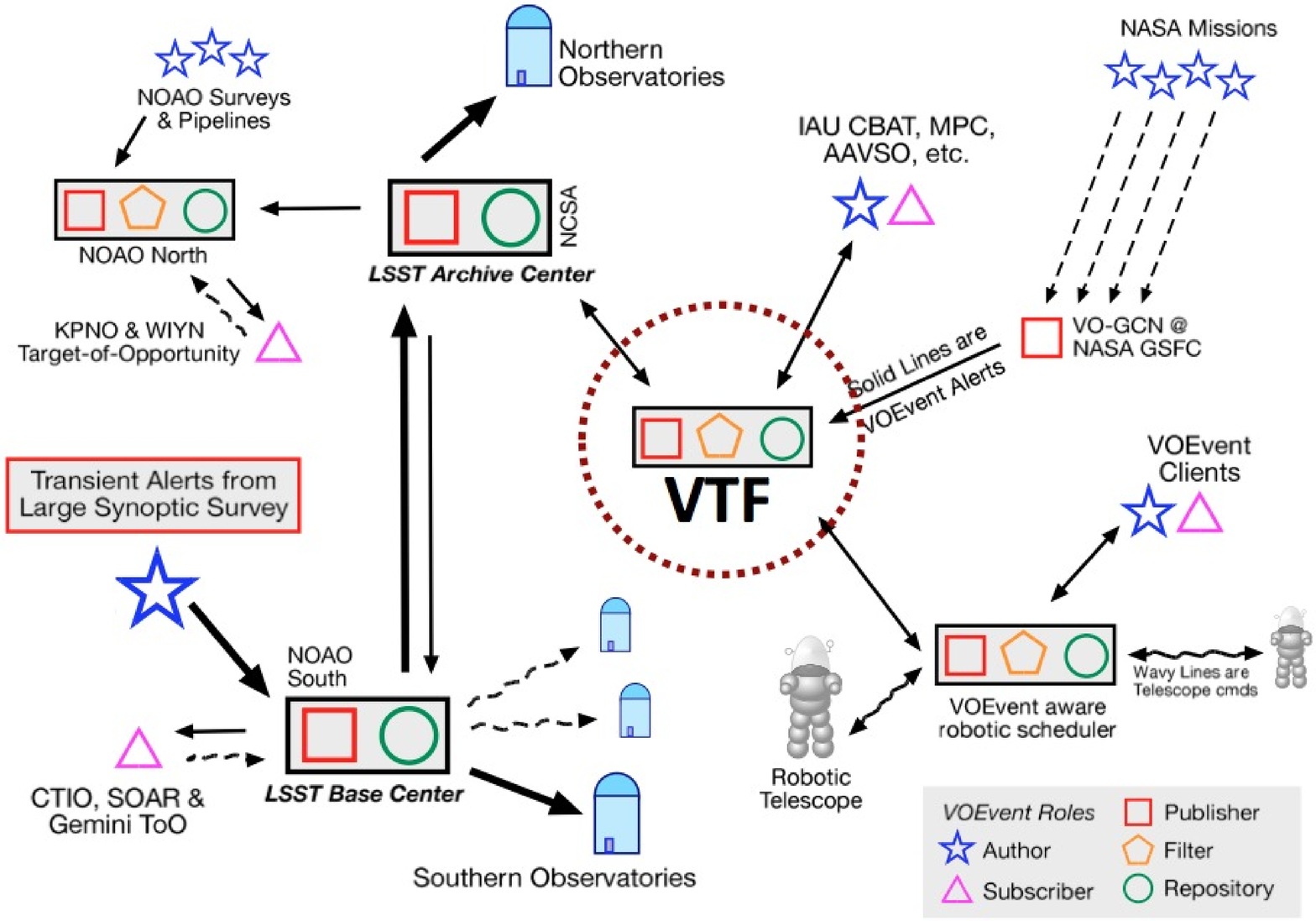} 
\caption{This shows a possible architecture for the dissemination of LSST events.}
\label{fig1}
\end{center}
\end{figure}

The current event infrastructure is an ad hoc arrangement mixing unwritten agreements, ideology and hacks that work. There are no inherent conceptual issues with this framework (see Fig.~\ref{fig1} for how LSST events could be disseminated by it) but with its increasing uptake and usage by new projects, its various components need to be standardized to ensure interoperability. Specifically, the format for describing infrastructure metadata, a protocol for accessing and querying stored event notifications, and the transport mechanisms used to distribute event notifications need to be specified.

\section{Recommendations and schedule}
The VAO Transient Facility can meet the specific needs of the transient community in terms of event notification and brokering capabilities by offering services based on the SkyAlert system and ensuring that the event infrastructure is completely formalized. The development schedule for such a venture is driven by consideration of a number of potential external science collaborations, specifically the LSST 2012 Data Challenge (June 2012), in which event notifications will be circulated at a ~25\% expected data rate, and the GAIA Event Challenge (March 2013). Obviously such activities would serve as a good test and demonstrator of the capabilities of the facility. To meet these, we propose to have a cluster-based fully-tested installation by mid-late 2012 compliant with a fully-specified event infrastructure.

\section*{Acknowledgements}
This work has been funded through NSF grants AST-0834235, AST-0909182 and OCI-0915473.

\end{document}